\documentclass[a4paper]{jpconf}
\usepackage{graphicx}
\begin{document}
\title{Isolation of gravitational waves from displacement noise and utility of a time-delay device}

\author{Kentaro Somiya$^1$, Keisuke Goda$^2$, Yanbei Chen$^1$, and Eugeniy E. Mikhailov$^3$}

\address{$^1$ Max-Planck Institut f\"ur Gravitationsphysik, Am M\"uhlenberg 1, 14476 Potsdam Germany}
\address{$^2$ LIGO Laboratory, Massachusetts Institute of Technology, Cambridge, MA 02139, USA}
\address{$^3$ Department of Physics, College of William \& Mary, Williamsburg, VA 23187, USA}

\ead{somiya@aei.mpg.de}

\begin{abstract}
Interferometers with kilometer-scale arms have been built for gravitational-wave detections on the ground; ones with much longer arms are being planned for space-based detection.  One fundamental motivation for long baseline interferometry is from displacement noise. In general, the longer the arm length $L$, the larger the motion the gravitational-wave induces on the test masses, until $L$ becomes comparable to the gravitational wavelength. Recently, schemes have been invented, in which displacement noise can be evaded by employing differences between the influence of test-mass motions and that of gravitational waves on light propagation.  However, in these schemes, such differences only becomes significant when $L$ approaches the gravitational wavelength. In this paper, we explore a use of artificial time delay in displacement-noise-free interferometers, which will shift the frequency band of the effect being significant and may improve their shot-noise-limited sensitivity at low frequencies. 

\end{abstract}

\section{Introduction}

Gravitational waves are ripples of space-time curvature caused by violent astronomical events such as supernova explosion or binary-stars coalescence.  The waves, upon arrival at the earth, are very weak.  For example, although currently operating gravitational-wave detectors can detect tiny strains of $h\sim 10^{-22}$ (at $\sim$100\,Hz), astrophysical estimates still predict a low chance for detection. Nevertheless, recent and projected improvements in sensitivity may soon allow us to have sizable chances of detecting gravitational waves. 

\begin{figure}[h]
\begin{minipage}{14pc}
 \includegraphics*[height=4.25cm]{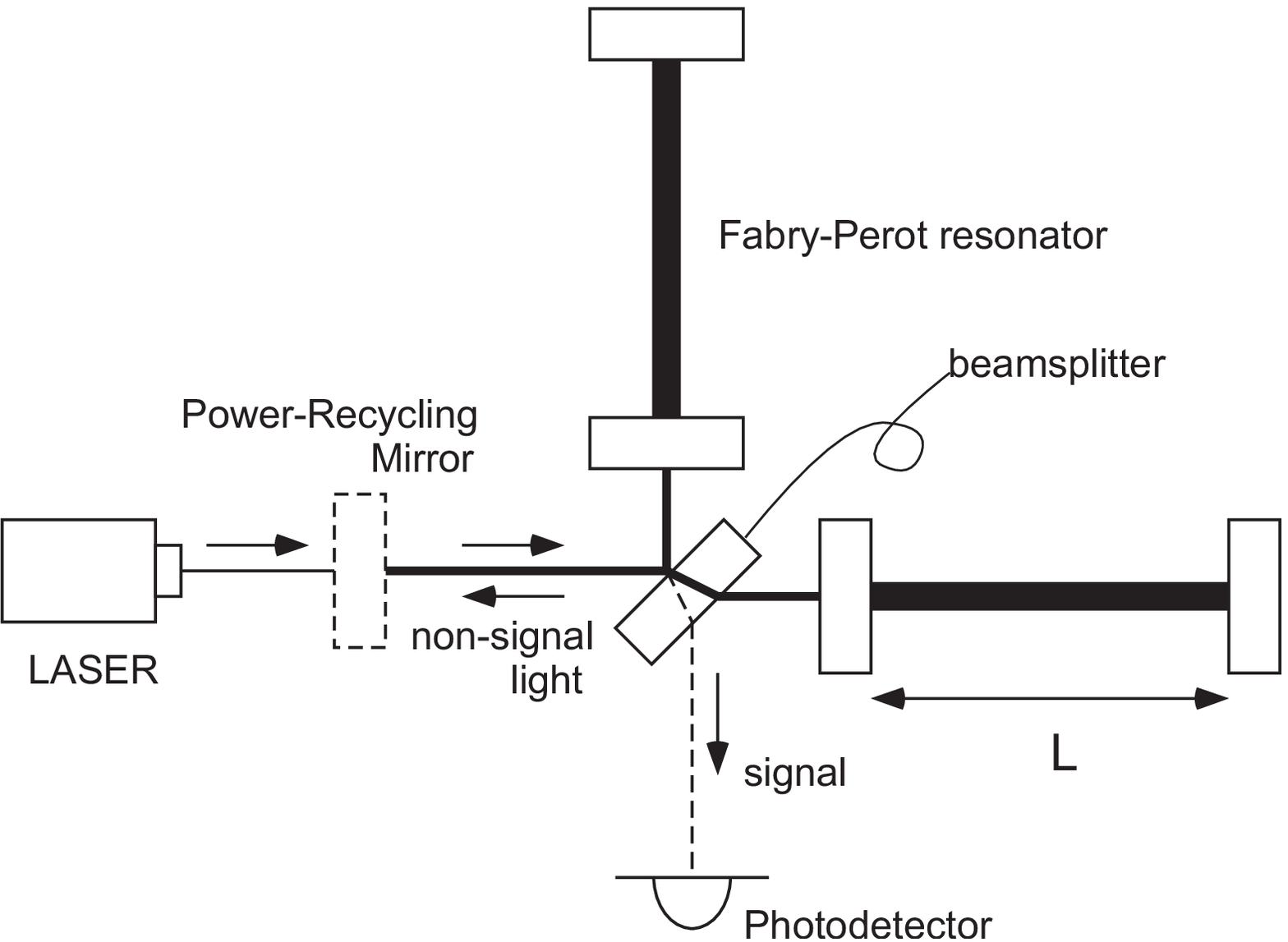}
 \caption{Fabry-Perot Michelson interferometer (FPMI) with a power-recycling cavity.\label{fig:FPMI}}
\end{minipage}\hspace{2pc}%
\begin{minipage}{22pc}
 \includegraphics*[height=4.2cm]{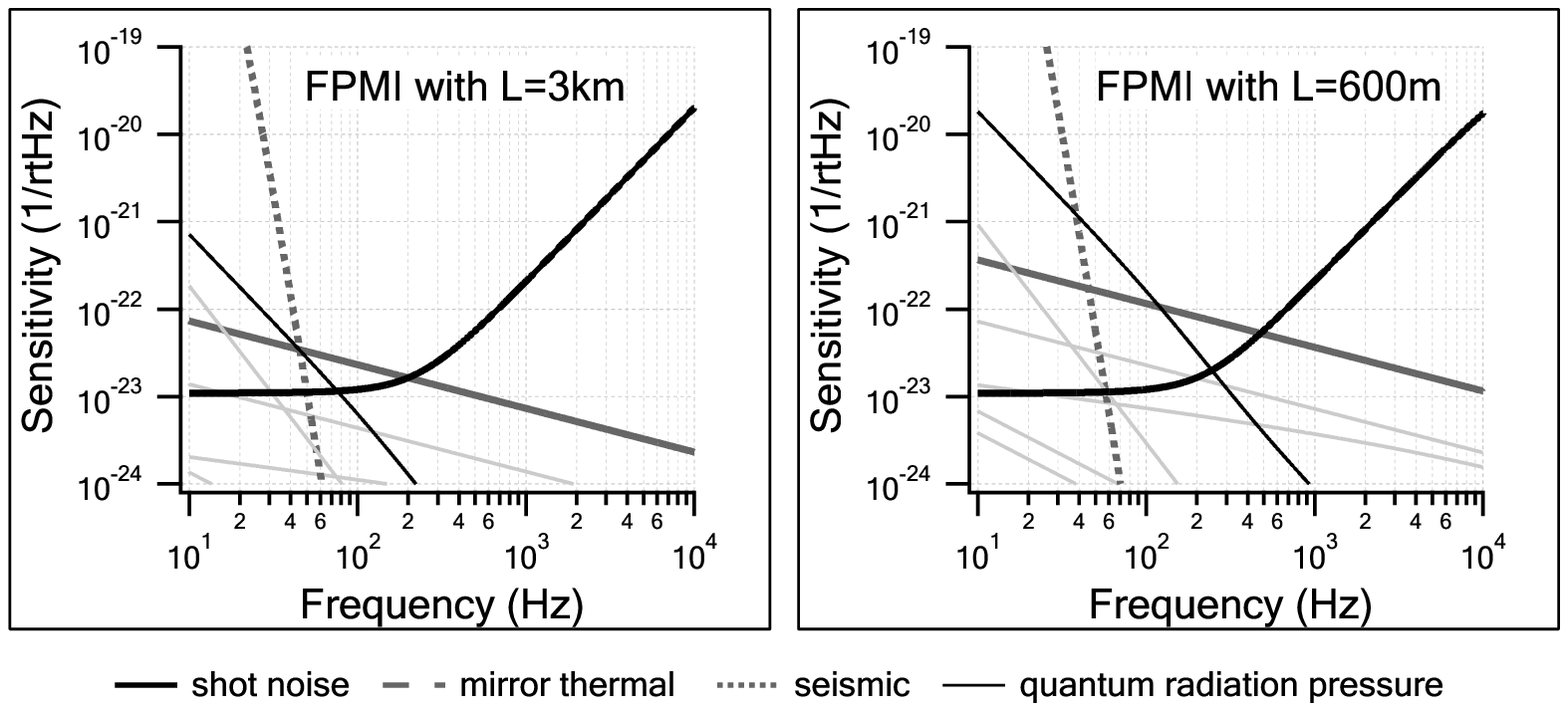}
 \caption{Calculated noise spectrum of FPMI with the arm length of 3 km and 600 m that have the same level of shot noise.\label{fig:3kAnd600m}}
\end{minipage} 
\end{figure}

The basic design of current gravitational-wave detectors~{\cite{LIGO}\cite{VIRGO}\cite{GEO}\cite{TAMA}} is based on Michelson interferometry. The laser beam splitted to two long arms senses a change of space-time during the round trip, being reflected back by mirrors at the end of the arms. A few more mirrors are added to the Michelson interferometer to make resonant cavities that can enhance the laser power and/or the gravitational-wave signals (see Fig.~\ref{fig:FPMI}).

Sensitivity of a detector is determined by the ratio of the signal and noise at each frequency. There are various kinds of noise on the detector, among which the most fundamental one is shot noise that comes from vacuum fluctuation of the light field. Shot-noise-limited sensitivity can be improved with an increase of the signal by making arms longer or by raising the light power in the arms. Recent improvement in development of the high-stability laser allows us to assume more than 100~W light to be injected to the interferometer. Inside the arm cavities could be more than few mega-watts with a use of low-loss mirrors.

However, we cannot improve the sensitivity of a gravitational-wave detector only by increasing the power. Besides shot noise, displacement noise due to thermal fluctuation or seismic motion limits the sensitivity, and displacement noise increases with the gravitational-wave signal when we raise the light power. Figure~\ref{fig:3kAnd600m} shows noise spectra of Fabry-Perot Michelson interferometers with the arm length of 3~km and 600~m. The laser power circulating in the arm cavities is 5 times higher in the 600~m interferometer so that the shot noise level appears to be same as the 3~km interferometer. One can see displacement noise is larger in the smaller interferometer, which can be explained in two ways: (i) the motion of the mirror in displacement ($\mathrm{m}/\sqrt{\mathrm{Hz}}$) is same and the sensitivity in strain ($1/\sqrt{\mathrm{Hz}}$) is derived from that divided by the arm length, or (ii) the detector is more sensitive to the motion of the mirror when it has more power inside the arm cavity. In this paper, we pick a sum of mirror thermal noise and seismic noise, shown in the graph with reasonable parameters currently available, as displacement noise for discussions, ignoring other small noise shown by light-gray lines and quantum radiation-pressure noise that seems to limit the sensitivity of 600~m interferometer because of the use of unnecessarily high power in the cavity.

Consequently gravitational-wave detectors should have as long arms as possible to decrease displacement noise appearing on the strain sensitivity. In other words, a detector could be shorter, which means cheaper, if it is possible to remove displacement noise; or we can say a detector could be sensitive at low frequencies with a same size.

Let us see if there is a way to remove displacement noise. First of all responses of a detector to gravitational waves and displacement noise must be different, otherwise the signal disappears with the removal of displacement noise. Indeed, they can be different.~\cite{Malik} One difference appears in a simple Michelson interferometer, where a gravitational-wave signal can flip the sign between the former half and the latter half of the journey at around a particular frequency and its odd harmonics. At the exact particular frequencies, displacement noise of the end mirrors remains and a gravitational-wave signal disappears. It is the other way round from what we want to do.

The key to eliminate displacement noise is to probe the mirror motion more than once. Probably the most common situation is with a Sagnac interferometer.~\cite{Sagnac} Splitted by a beamsplitter, two lights probe the motion of two end mirrors in a different order with some time delay. At particular frequencies depending on the time delay, we can remove displacement noise with gravitational-wave signals remaining. Also, recently Kawamura and Chen found a way to remove displacement noise at all the frequencies using a combination of multiple beams that probe each mirror motion more than once; the system is called displacement-noise-free interferometer (DFI).~{\cite{DFI1}\cite{DFI2}\cite{DFI3}} We shall name the former one a {\it narrow-band} DFI and the latter one a {\it broadband} DFI.

However, we still need long arms with DFI configurations. The frequency at which displacement noise can be cancelled in a Sagnac interferometer depends on the time difference between two probings, which means it is a very high frequency unless the arm is very long. The response of a broadband DFI to gravitational waves decays at low frequencies by $f^2$ or $f^3$, depending on the configuration, up to a certain frequency that is also determined by the arm length so that it also requires the long arms.

Here one might reconcile to make a sensitivity of a short-arm detector comparable to that of a long-arm detector, but we would like to point out there could be a way to circumvent the problem. The reason why we need long arms in a non-DFI detector is to receive gravitational waves without probing mirror motions, and the reason why we need long arms in a DFI detector is to wait for some time that makes the phase of gravitational waves change. The former reason can be answered only with long arms unless the carrier is changed from light to something slow~\cite{atomic}, while the latter one can be answered not only with long arms but also with a use of a time-delay device that can trap the light for some time. Note that the light being trapped by a time-delay device, with our definition, does not sense gravitational waves during the trapment. There are many ways to implement time delay to the light: optical fibers, multiple cavities, or a quantum optics device based on {\it electromagnetically induced transparency} (EIT).~\cite{EIT} 

In this paper, applying a time-delay device to a narrow-band and to a broadband DFI detector, we will see how the sensitivities can be improved at low frequencies. Section~\ref{sec:2} is a review of how to calculate a response to gravitational waves, followed with a couple of clues to isolate gravitational-wave signals from displacement noise. Section~\ref{sec:3} shows calculated sensitivity curves of a Sagnac interferometer with and without a time-delay device. Section~\ref{sec:4} shows sensitivities of a broadband DFI with and without time delay. In Sec.~\ref{sec:5}, we shall make discussions about fluctuation of the time delay that is anticipated to be additional noise. At last, Section~\ref{sec:6} is a summary.

\section{Response to gravitational waves}\label{sec:2}

Let us derive the response of an interferometer to gravitational waves. Orientation of the wave source is expressed with $\theta$ and $\psi$ by 
\begin{eqnarray}
\vec{e}_x&=&(\cos{\theta}\cos{\psi},\ \cos{\theta}\sin{\psi},\ -\sin{\theta})\ ,\\
\vec{e}_y&=&(-\sin{\psi},\ \cos{\psi},\ 0)\ ,\\
\vec{e}_z&=&(\sin{\theta}\cos{\psi},\ \sin{\theta}\sin{\psi},\ -\cos{\theta})\ .
\end{eqnarray}
We assume that the wave propagates along the $z$ axis. At the time when the waves and the light of an interferometer reach the same point, the speed of light slightly changes by $h[t]$. Integration of the effect over one-way journey on the straight arm starting from $\vec{r}$ at $t=t_1$ ending at $t=t_2$ gives the phase shift:
\begin{eqnarray}
\phi_\mathrm{GW}(\theta,\psi,t_1,t_2,\vec{r})=\omega_0\int_\mathrm{t_1}^\mathrm{t_2}h[t-{\vec{r}\cdot\vec{e}_z}/{c}-(t-t_1)\vec{n}\cdot\vec{e}_z]dt\ ,\label{eq:GWtime}
\end{eqnarray}
where $\omega_0$ is the frequency, $c$ is the speed, and $\vec{n}$ is a unit vector of the light. Gravitational waves have two polarizations, which are $+$ mode that causes a differential change between $x$ axis and $y$ axis, and $\times$ mode that causes a differential change between $x+y$ axis and $x-y$ axis. The following function will be multiplied to $\phi_\mathrm{GW}$ of each mode:
\begin{eqnarray}
F_+=(\vec{e}_x\cdot\vec{n})^2-(\vec{e}_y\cdot\vec{n})^2\ ,\ \ \ F_\times=2(\vec{e}_x\cdot\vec{n})(\vec{e}_y\cdot\vec{n})\ .
\end{eqnarray}
The phase shift $\phi_\mathrm{GW}$ is a function of $\theta$ and $\psi$. We can take an average all over the sky as
\begin{eqnarray}
\phi_\mathrm{GW}(t_1,t_2,\vec{r})=\frac{1}{\sqrt{4\pi}}\int_0^{2\pi}\int_0^{\pi/2}\phi_\mathrm{GW}(\theta,\psi,t_1,t_2,\vec{r})\sin{\theta}d\theta d\psi\ ,
\end{eqnarray}
which will be then Fourier transformed into $\tilde{\phi}_\mathrm{GW}(\omega)$ with $t_2-t_1$ and $\vec{r}$ being given.

In \eqref{eq:GWtime}, one can see a possible difference from the response to displacement noise. There is a factor $(1-\vec{n}\cdot\vec{e}_z)$ in front of $t$ inside the bracket, which shows that the light senses gravitational waves with a Doppler shift when the waves do not come from the top of the sky. If one can somehow prepare a situation where displacement noise at some frequencies disappears, the detector can sense waves whose frequency is shifted by this effect without sensing displacement noise. Note that, if the light returns to the same direction as its coming journey, the Doppler shift cancels after the round trip. We can have one direction for the coming beam, another direction for the going beam, and a mirror motion is measured in the bisector angle of these two directions. We can say this is a two-dimensional measurement, comparing with a usual one-dimensional measurement, in which the light is normally incident on a mirror.

\begin{figure}[h]
\begin{center}
 \includegraphics*[height=3.5cm]{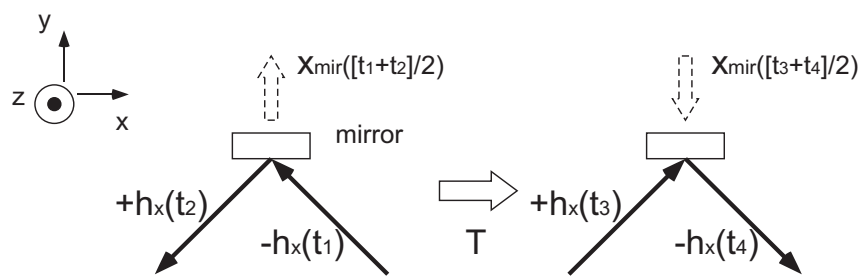}
 \caption{Displacement noise disappears at a particular frequency while gravitational-wave signals remain in a two-dimensional measurement.\label{fig:2Dmeasurement}}
\end{center}
\end{figure}

Another possible difference from the response to displacement noise also appears at the two-dimensional measurement. See Fig.~\ref{fig:2Dmeasurement}. Assuming the orientation of gravitational waves is that with $\theta=\psi=0$, and the incident angle of the light is 45 degrees, we have $F_\times=+1$ for the outgoing beam to and the incoming beam from the left side ($h_\times$ at $t=t_2$ and $t_3$), and $F_\times=-1$ for the beam to and from the right side ($h_\times$ at $t=t_1$ and $t_4$); $F_+$ is always zero. If the phase shift after a time interval $T$ is $\pi$, and we add the results before and after the interval, displacement noise will disappear. On the other hand, gravitational-wave signal will remain due to the flip of the sign of $F_\times$ between $t_1$ and $t_3$, or $t_2$ and $t_4$, which cancels the phase shift of $h_\times$ after the time interval. If the incident angle is 0 degrees, $F_\times$ is always zero and $F_+$ is always $-1$, so the phase shift after time interval makes the cancellation of displacement noise and gravitational waves occur at the same time.

\section{Sagnac interferometer as a narrow-band DFI with/without time delay}\label{sec:3}

\subsection{Cancellation of displacement noise}
\begin{figure}[h]
\begin{minipage}{16pc}
\begin{center}
 \includegraphics*[height=8cm]{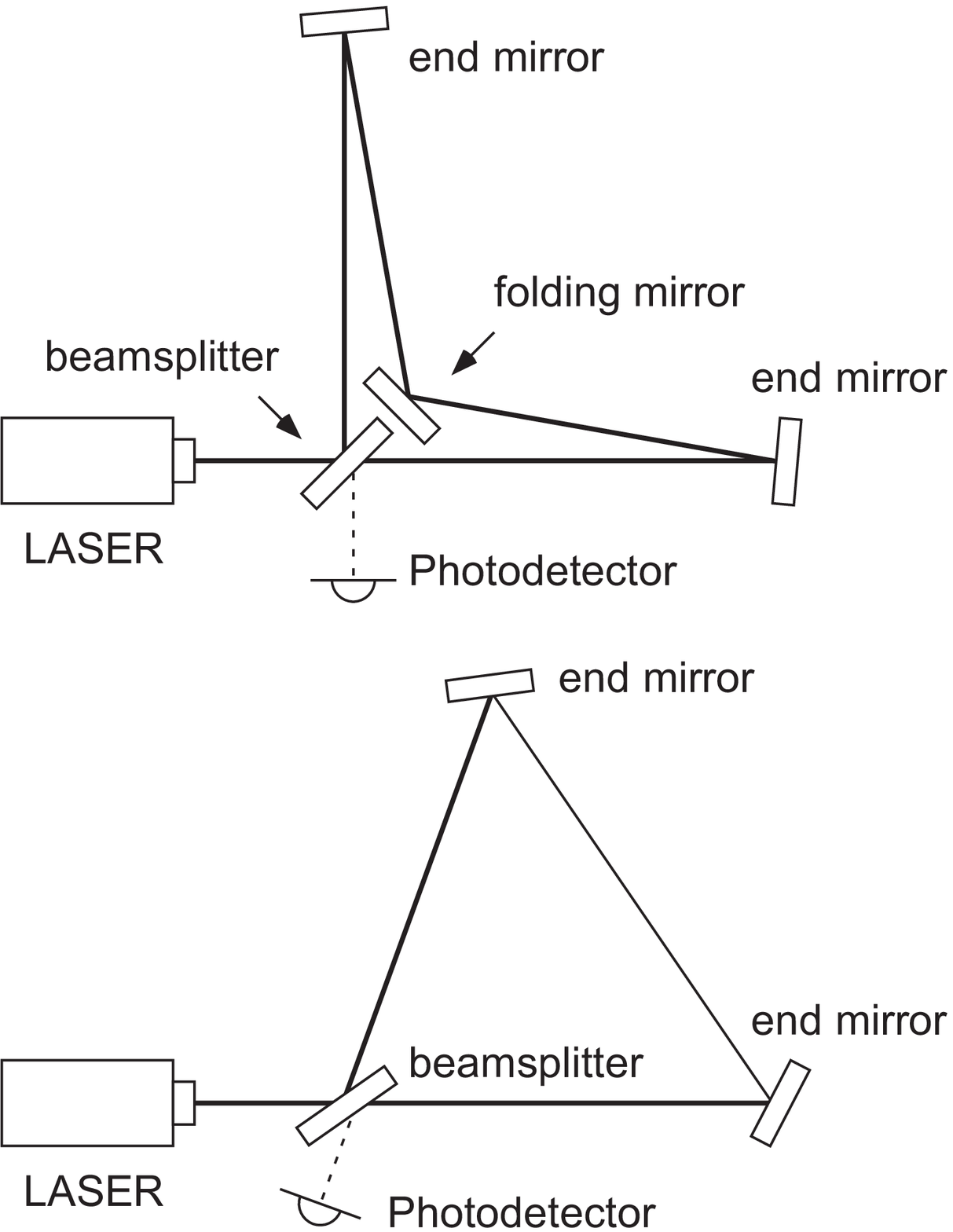}
 \caption{A folded Sagnac interferometer (top) and a triangle Sagnac interferometer (bottom).\label{fig:Sagnacs}}
\end{center}
\end{minipage}\hspace{2pc}%
\begin{minipage}{20pc}
\begin{center}
 \includegraphics*[height=8cm]{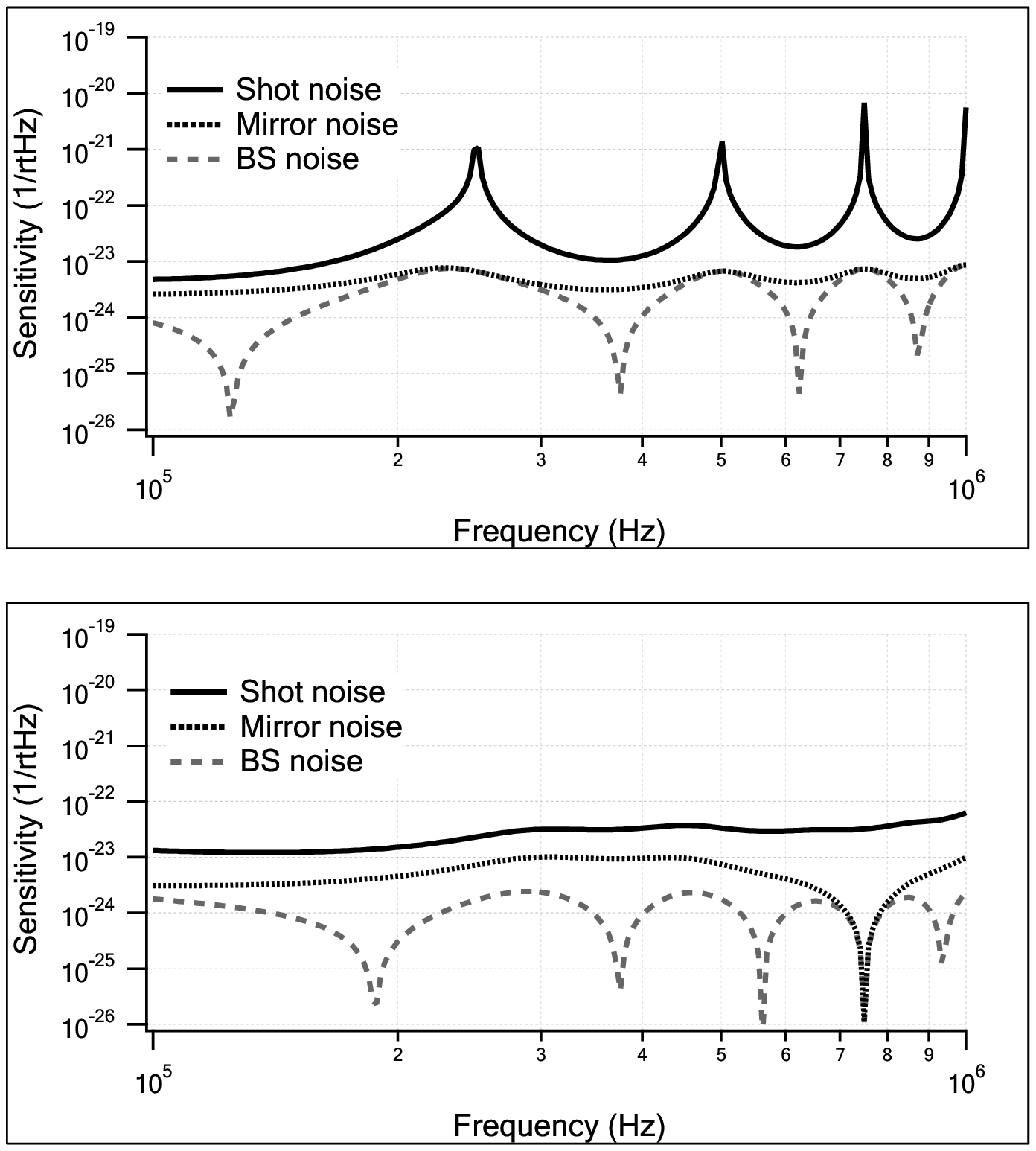}
 \caption{Shot noise and displacement noise of the folded Sagnac interferometer (top) and the triangle Sagnac interferometer (bottom).\label{fig:nonTD2}}
\end{center}
\end{minipage} 
\end{figure}

We shall see the noise-cancellation effect in a Sagnac interferometer, for example. Figure~\ref{fig:Sagnacs} shows a couple of Sagnac configurations and their noise spectra. In one Sagnac, the beams propagating toward two directions after a beamsplitter get reflected by end mirrors almost to the same way, reaching a mirror placed close to the beamsplitter that folds the beams to the end mirrors on the other side, and then two beams interfere to send only differential signals to a photodetector placed at the other side from the laser. The motion of each end mirror is probed by the light with normal incident, so that this is a one-dimensional measurement. There is no way to extract gravitational-wave signals without displacement noise at any frequency. In the other Sagnac, the beams reflected by end mirrors directly go to the other end mirrors. The direction of the mirror motion that is measured and the direction of the light that senses gravitational waves are different. This is a two-dimensional measurement.

Besides the motion of the end mirrors, we should take a motion of the beamsplitter into account. To eliminate displacement noise of the mirrors and the beamsplitter at a same frequency, the distances between the optical components should be chosen appropriately. It has turned out from our calculation that the frequencies of the cancellation cannot be same if the configuration is in a regular triangle. We choose the distance between one of the mirrors and the beamsplitter to be 600~m, and the distance between the two mirrors to be 400~m, which makes the cancellation frequencies exactly equal as is shown in Fig.~\ref{fig:nonTD2} (bottom panel). The laser power injected to the Sagnac interferometer is 100~MW. The same sensitivity can be realized at low frequencies by 100~W laser if one uses power-recycling~\cite{PR} and signal-recycling~\cite{SR} techniques, with each recycling gain to be 1000. The power recycling will be realized by a mirror that reflects back the light returned towards the laser (bright port), and the signal recycling will be realized by a mirror at the other side of the power-recycling mirror (dark port) to reflect back the signal components. At frequencies higher than $\sim200$~Hz, in the case shown above, the signal starts cancellation during the circulation in the signal-recycling cavity. Here we assume no recycling just for simplicity. Orientation of gravitational waves is averaged all over the sky. 


\subsection{Implementation of a time-delay device}

It is good to have a sensitivity not limited by displacement noise, but the displacement-noise-free frequency is actually higher than the frequencies where displacement noise limits the sensitivity in a usual detector (Fig.~\ref{fig:3kAnd600m}), which means it is quite useless. Now we shall put a time-delay device into this Sagnac interferometer. 

\begin{figure}[h]
\begin{minipage}{16pc}
\begin{center}
 \includegraphics*[height=4cm]{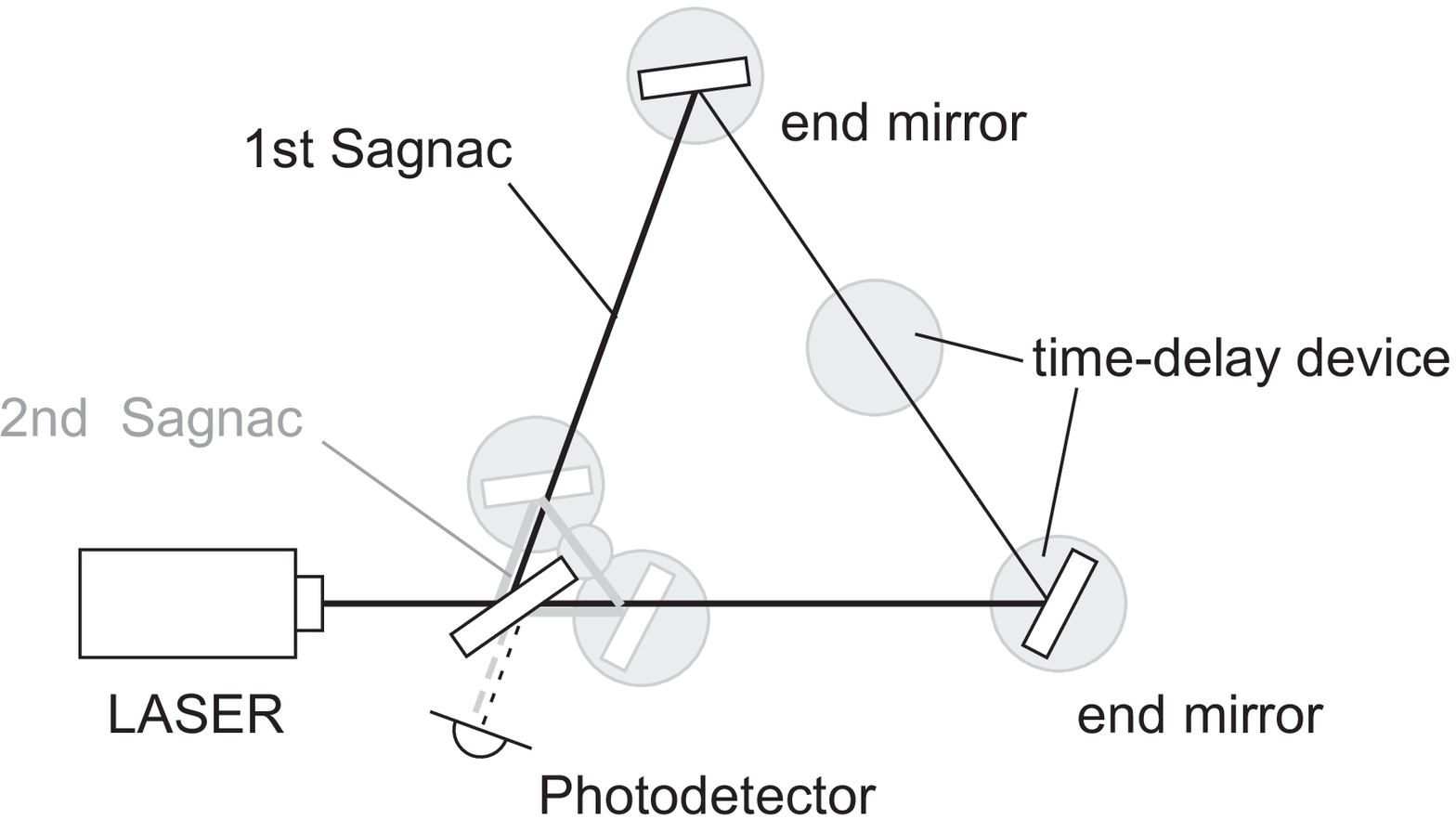}
 \caption{Sagnac interferometer with time-delay devices.\label{fig:TDSagnac}}
\end{center}
\end{minipage}\hspace{2pc}%
\begin{minipage}{20pc}
\begin{center}
 \includegraphics*[height=4cm]{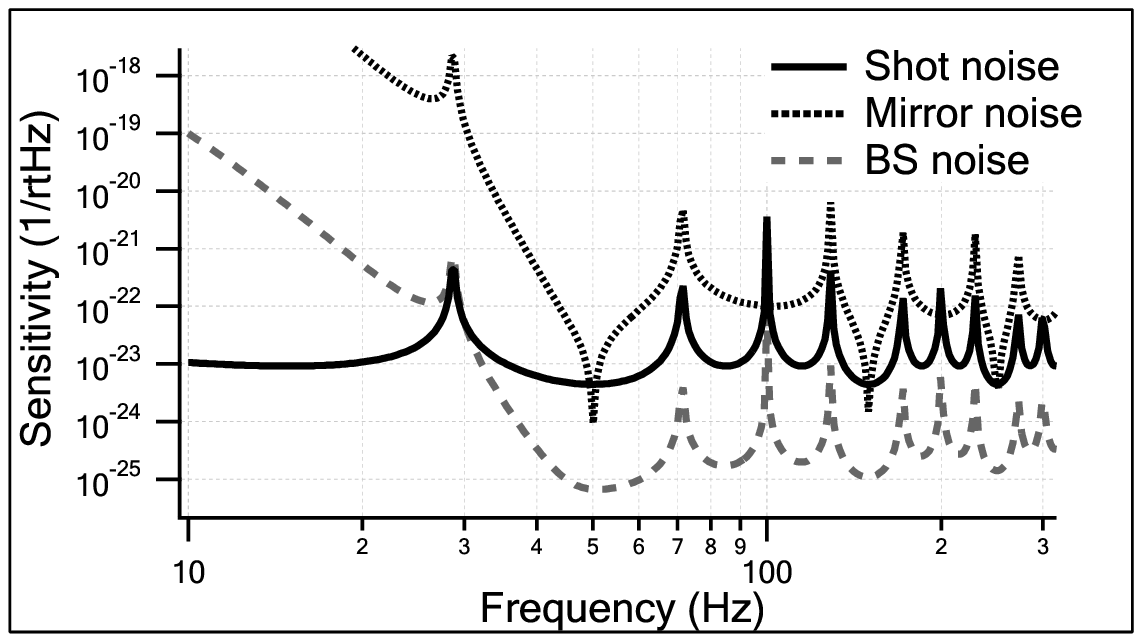}
 \caption{Sensitivity of the time-delay Sagnac interferometer.\label{fig:withTD}}
\end{center}
\end{minipage} 
\end{figure}

Let us first ignore noise of a time-delay device and see if there is a way to cancel displacement noise of all the optical components. Implementing time-delay devices at the corners and the middle point between two mirrors (Fig.~\ref{fig:TDSagnac}), we can make a displacement-noise-free frequency for the mirrors different from the frequency where a gravitational-wave signal cancels, which appears as the peak of the shot-noise curve (Fig.~\ref{fig:withTD}). However it has turned out that a displacement-noise-free frequency for the beamsplitter is always same as the gravitational-wave-cancellation frequency, so we need another Sagnac interferometer that has the beamsplitter commonly used as is shown in Fig.~\ref{fig:TDSagnac}. The outputs of the first Sagnac that senses gravitational waves and the second Sagnac that does not sense the waves but does probe the motion of another pair of mirrors and the common beamsplitter will be subtracted to cancel displacement noise of the beamsplitter. Here the time delay is set to 10~ms each. While the same level of displacement noise is used, we can see a big improvement in the sensitivity at around 50~Hz compared with Fig.~\ref{fig:3kAnd600m}.

\section{Broadband DFI with/without time delay}\label{sec:4}

\begin{figure}[h]
\begin{minipage}{16pc}
\begin{center}
 \includegraphics*[height=6cm]{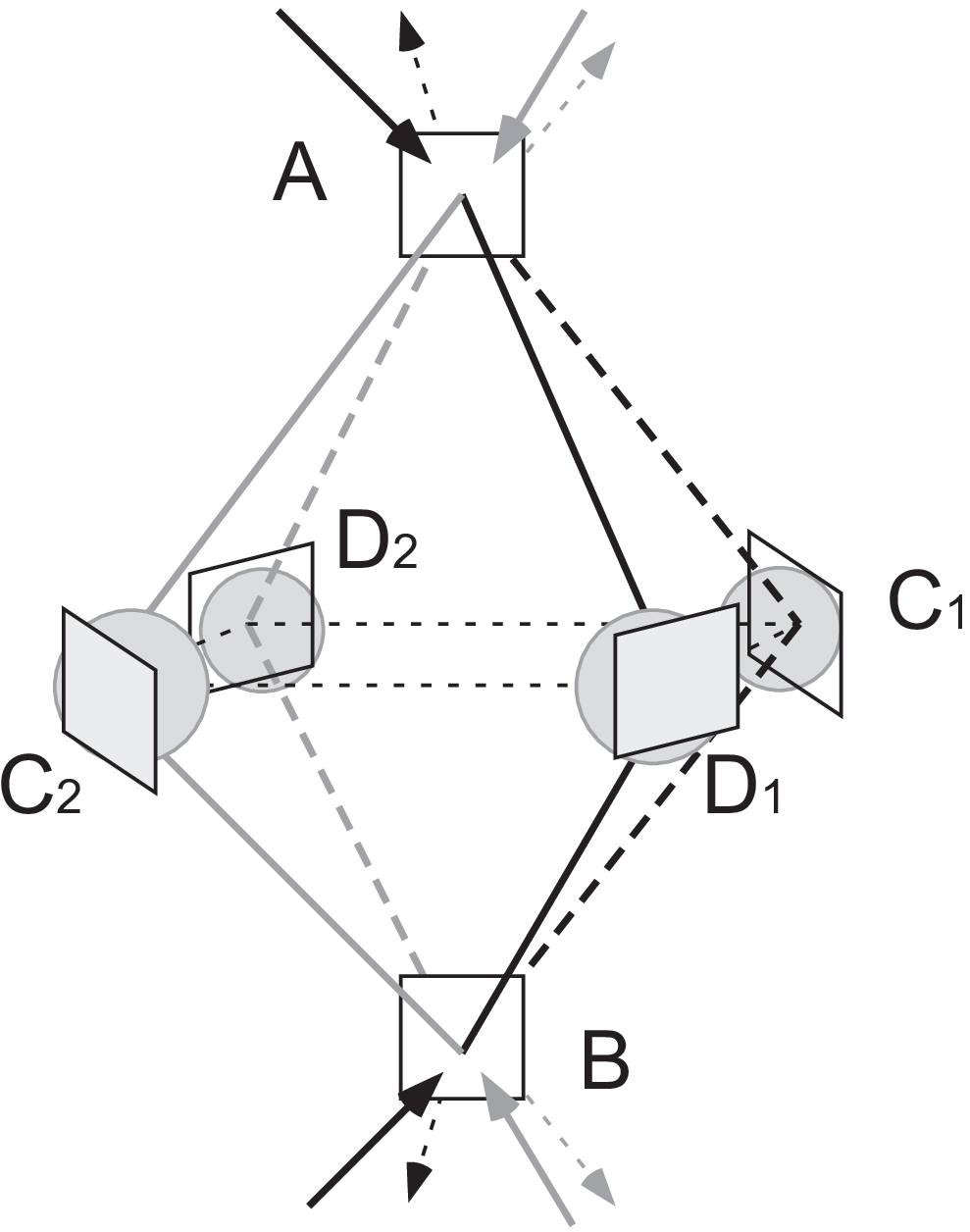}
 \caption{A 3D Mach-Zehnder interferometer as a broadband DFI.\label{fig:3DMZ}}
\end{center}
\end{minipage}\hspace{2pc}%
\begin{minipage}{20pc}
\begin{center}
 \includegraphics*[height=4cm]{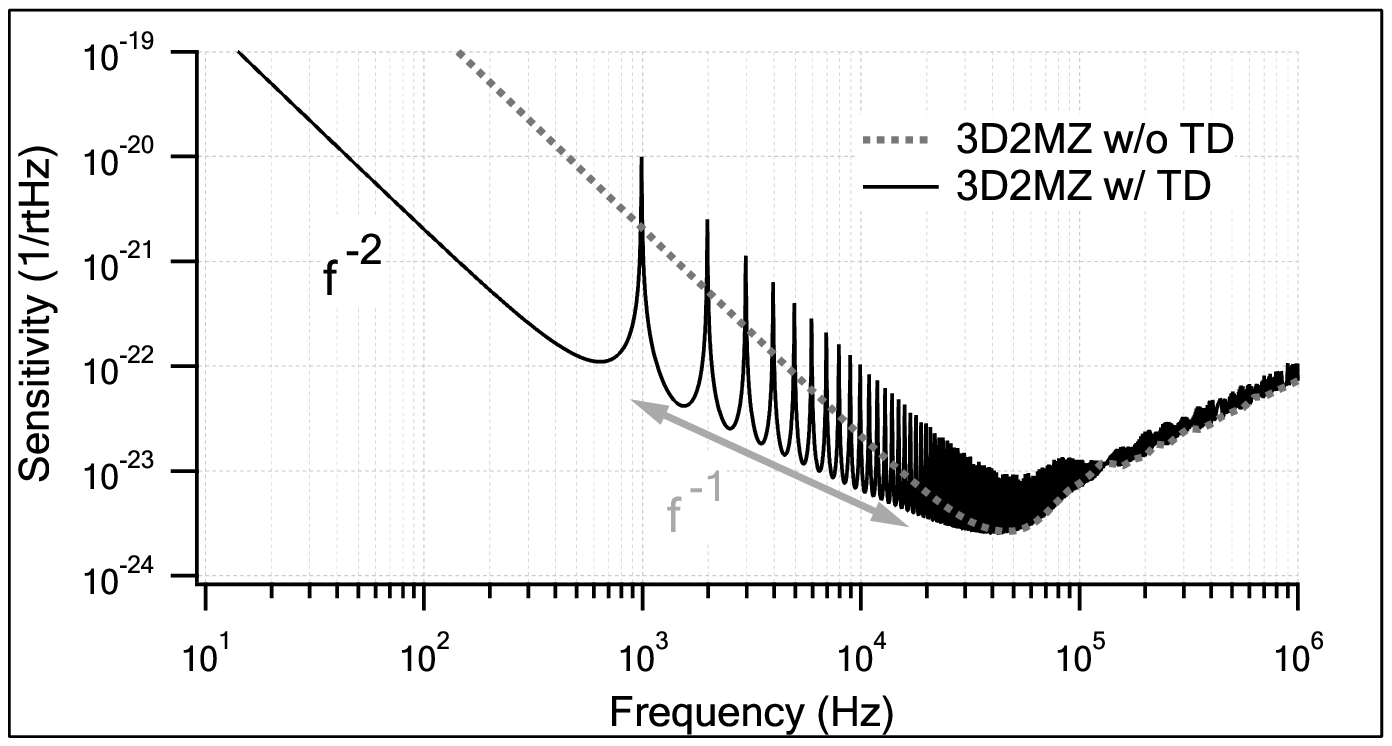}
 \caption{Sensitivity of a 3D Mach-Zehnder interferometer with and without time delay. Provided displacement noise be perfectly cancelled, it is all shot noise that makes the sensitivity curve. Here orientation of gravitational-wave source is averaged all over the sky.\label{fig:3D2MZsens}}
\end{center}
\end{minipage} 
\end{figure}

In a narrow-band DFI detector, displacement noise is cancelled at a particular frequency as the motion of the mirror is probed twice with a time difference. The key idea of a broadband DFI is to use more than one beam and to probe each motion of all the mirrors at a same time with different beams. Figure~\ref{fig:3DMZ} shows 3-dimensional (3D) Mach-Zehnder interferometer that works as a broadband DFI.~\cite{DFI3} There are 8 beams inside the octahedron. One beam starting from $A$ and reaching $B$ via $C_1$ or $D_1$ carries displacement noise of $C_1$ or $D_1$ that can be cancelled by taking subtraction with what is carried by the beam starting from $B$ and reaching $A$ via $C_1$ or $D_1$ while displacement noise of $A$ and $B$ are not cancelled by this subtraction. Another pair of beams probing $C_2$ or $D_2$ can be also free from displacement noise of the mirrors and carry displacement noise of $A$ and $B$. Taking subtraction of these two sets that probe displacement noise of the beamsplitters from different sides, one can then cancel all displacement noise. Difference of displacement noise and gravitational-wave signal appears at high frequencies. After the subtraction, the response to gravitational waves shows an $f^2$ curve in the 3D configuration, which means the shot-noise-limited sensitivity spectrum becomes $f^{-2}$, up to a frequency determined by the arm length:
\begin{eqnarray}
f_\mathrm{cutoff}=c/(2L)\ .
\end{eqnarray}
The dotted curve in Fig.~\ref{fig:3D2MZsens} shows the sensitivity of the 3D Mach-Zehnder interferometer. Each arm is 3~km and the circulating power is 100 MW without signal-recycling. The cutoff frequency is then $f_\mathrm{cutoff}=500$~kHz, below which the sensitivity goes down by $f^{-2}$. The power recycling of Mach-Zehnder interferometer will be realized by installing one mirror between the laser and the interferometer and another mirror at the transmission port of the light to make a cavity. Signal recycling can be realized as well by installing a mirror at the other side of the beamsplitter after the injection of the laser and another mirror at the signal-extraction port to make another cavity. 

Now let us put time delay into this 3D Mach Zehnder interferometer. Although two light fields sense fluctuation of the same time-delay device, differential components will appear since the fields enter the device from opposite directions. Here it is just ignored.

The solid curve in Fig.~\ref{fig:3D2MZsens} shows the sensitivity. The time delay is set to $\tau=1$~ms. The slope of the sensitivity curve is $f^{-2}$ below $1/\tau$. There are peaks at the frequency and its harmonics since the phase shift in the time-delay device is $2\pi$ and the effect cancels. Except for the frequencies around the peaks, the sensitivity shows an $f^{-1}$ curve between $1/\tau$ and $f_\mathrm{cutoff}$. The difference of the slope in this region makes the sensitivity better at low frequencies compared with the same configuration without time delay. Note that the source orientation is averaged all over the sky. The peaks in the region of the time delay being effective are not shaved off by taking average since orientation has nothing to do during the time when the light is trapped in the time-delay device.

\section{Noise cancellation of a time-delay device}\label{sec:5}

So far, we have not taken noise of a time-delay device into account. Either an optical fiber or a Fabry-Perot cavity will impose much larger displacement noise than a single mirror. According to our calculation, shown in \ref{app}, EIT will impose noise much larger than the shot-noise level of a detector unless the device is cooled down to lower than a hundred mK or the size of the beam inside the device is increased as much as 1~m in diameter. While we shall keep looking for a way to reduce time-delay noise, which could hopefully be possible in a future, we can also try a configuration that is free from time-delay noise.

At each point where we put time delay, a pair of beams come from different directions. One of the light experiences time delay of one side of the device first and the other light experiences the other side first. Then both of them hit the mirror, reflected to the other side that the other light has experienced already. Time-delay noise appears due to the fact that two beams experience a same part of the device but in a different order. 

\begin{figure}[htbp]
\begin{center}
 \includegraphics*[height=3cm]{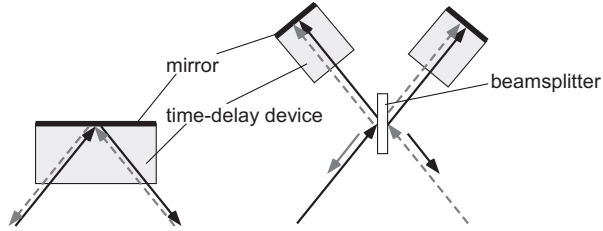}
 \caption{Time delay with counter-propagating beams in a single device (left panel) and with beams propagating in a pair of common device after a beamsplitter (right panel; a time-delay Michelson interferometer).\label{fig:EIT-MI}}
\end{center}
\end{figure}

We introduce a way to cancel time-delay noise by using a Michelson interferometer. See Fig~\ref{fig:EIT-MI}. The beams that come from different directions combine at a beamsplitter and make two beams. Each beam enters into a time-delay device that has a same level of the time delay effect with some fluctuations. Beams are recombined at the beamsplitter and the light goes to the other side from the injection with appropriate control of the Michelson interferometer. Both beams going to the other way after the Michelson interferometer has the same amount of the time-delay fluctuation that will be cancelled at the interference, or subtraction, at the main beamsplitter of the Sagnac interferometer.

\begin{figure}[h]
\begin{minipage}{16pc}
\begin{center}
 \includegraphics*[height=6cm]{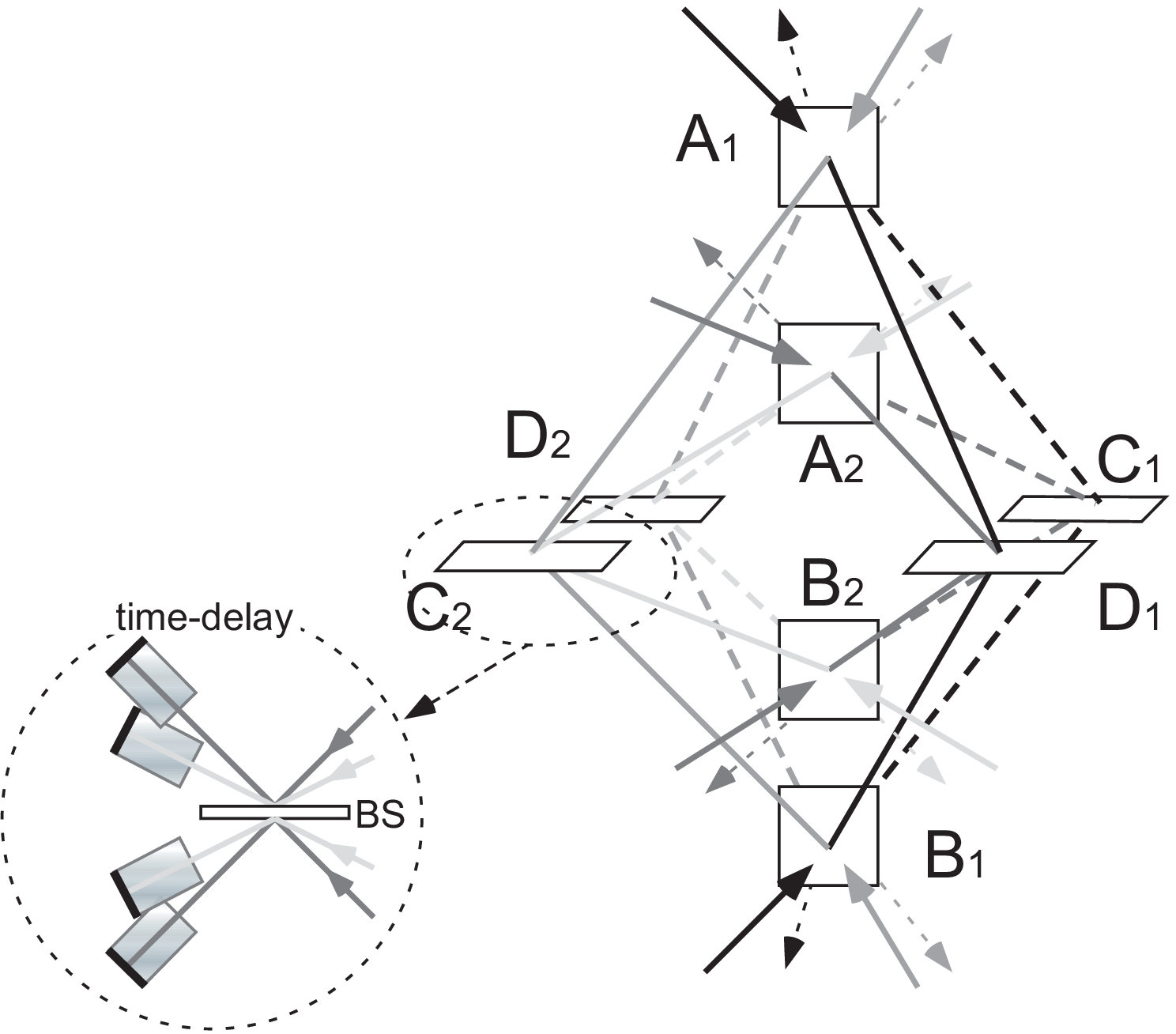}
 \caption{A 3D 4-Mach-Zehnder DFI.\label{fig:3D4MZ}}
\end{center}
\end{minipage}\hspace{2pc}%
\begin{minipage}{20pc}
\begin{center}
 \includegraphics*[height=5cm]{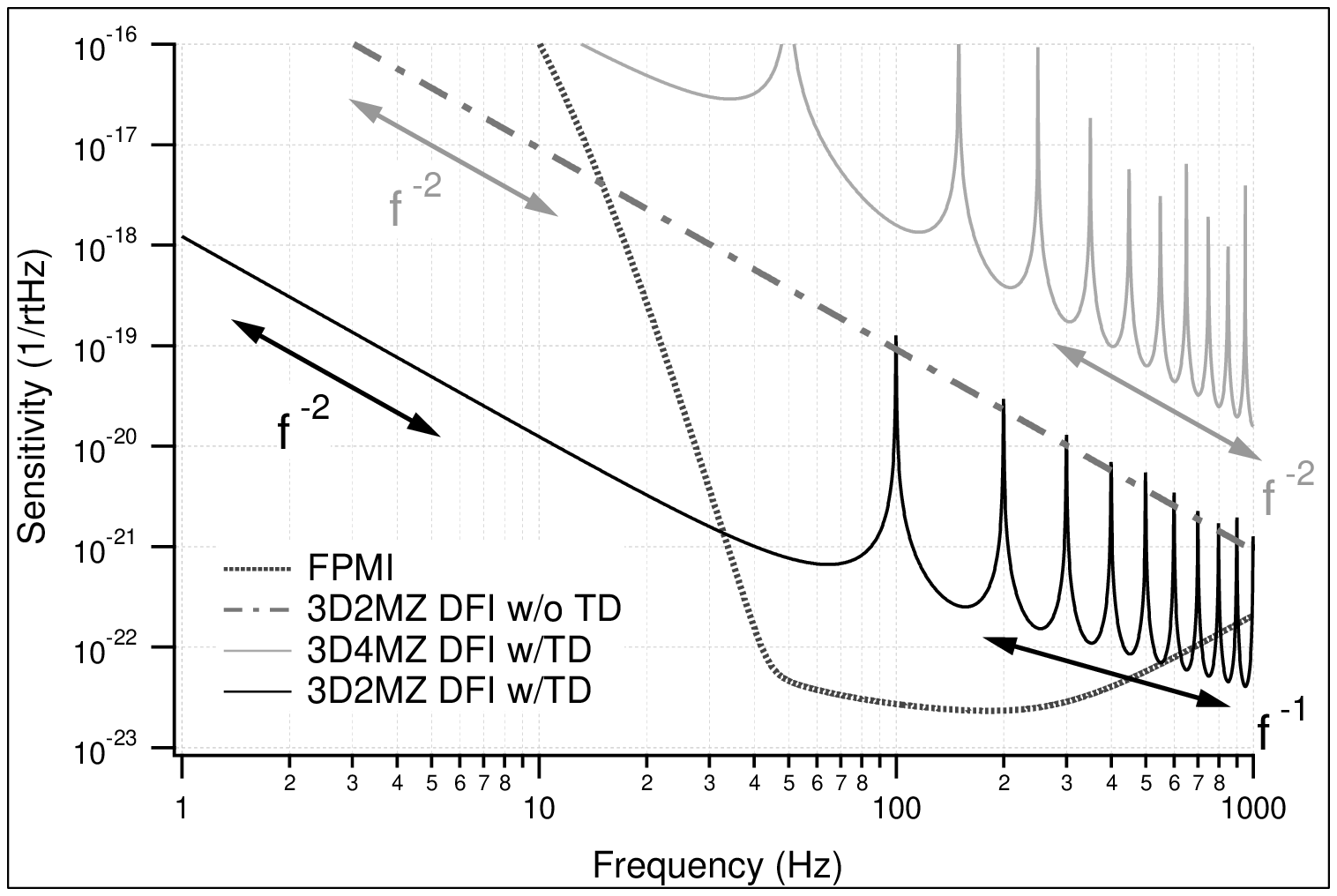}
 \caption{Sensitivity of a 3D 4-Mach-Zehnder interferometer compared with a conventional detector.\label{fig:3D4MZsens}}
\end{center}
\end{minipage} 
\end{figure}

However, displacement noise of the beamsplitter of this time-delay Michelson interferometer will be imposed to the system. Each beam probed the motion of this beamsplitter once from one side and once again after time delay from the other side. At the frequency such that the time delay makes the phase shift of the time delay exactly $2\pi$, displacement noise of the motion cancels out, but it means the time delay is not effective at the frequency. Displacement noise of the beamsplitter is unavoidable with the time-delay effect. In a Sagnac configuration, we have not found a way to remove this. Indeed, there are too many time-delay devices used in the Sagnac interferometer shown in Fig.~\ref{fig:TDSagnac}. In a broadband DFI, we can try to cancel beamsplitter noise of this time-delay Michelson interferometer by implementing another octahedron that shares the motion of the beamsplitters (Fig.~\ref{fig:3D4MZ}). Each beamsplitter at $C_1$, $C_2$, $D_1$ ,or $D_2$ that makes the time-delay Michelson interferometer is probed by two different light fields; one from the outer octahedron and the other from the inner octahedron. Appropriate combination of these two outputs, each of which is apparently free from the motion of other mirrors such as $A_1$ and $B_1$ or $A_2$ or $B_2$, eliminates displacement noise of the beamsplitters.

Now we have a detector that is free from laser noise, displacement noise of all the components, and time-delay noise. However, as is shown by a gray solid curve in Fig.~\ref{fig:3D4MZsens}, the sensitivity becomes worse. The slope is no more $f^{-1}$ like the 3D 2-Mach-Zehnder interferometer with time delay at any frequencies but $f^{-2}$ like a 3D DFI without time delay. The sensitivity is even worse than that of a 3D DFI without time delay probably because gravitational-wave signals cancel between two octahedrons. Here the time delay is $\tau=100$~ms and the incident power to the interferometer is 100~MW ($\sqrt{2}$ times larger power should be injected to the inner octahedron of 3D 4-Mach-Zehnder interferometer). The orientation of gravitational waves is fixed to the top of the sky (from $A$ to $B$) in order to compare the sensitivity with the conventional configuration that possesses the same arm length. Note that the sensitivity of a 3D 2-Mach-Zehnder DFI is better than that with the conventional configuration at frequencies lower than $\sim30$~Hz.

\section{Summary and discussions}\label{sec:6}

In this paper, we showed the possibility of using time-delay devices to improve the sensitivity of narrow-band and broadband displacement-noise-free gravitational-wave detectors --- even when these detectors have arms much shorter than the wavelength. A key to isolate gravitational-wave signals from displacement noise is the employment of a two-dimensional measurement, in which motions of mirrors are probed redundantly, by beams that propagate on different planes. While the difference between gravitational-wave signal and displacement noise becomes smaller at frequencies lower than $c/L$, this can be partially compensated by using time delay.

A narrow-band DFI like Sagnac interferometer has been known but not regarded to be useful for gravitational-wave detection since a noise-free frequency is so high that displacement noise is not a big problem. We showed that the frequency can be shifted low with time delay so as to make the effect be useful. Besides, the study of the narrow-band DFI has helped deep understanding of the noise-free effect. A broadband DFI that has been recently invented has a problem that the shot-noise level at low frequencies shows an $f^{-2}$ curve, at best, and the sensitivity is hardly better than that with the conventional configuration unless the input power be extremely high. We showed that the slope can be $f^{-1}$ with time delay and the sensitivity can be better than that with the conventional configuration at low frequencies.

A problem in the use of time delay is its fluctuation. In the case of EIT medium, noise caused by the thermal motion of molecules is overwhelming, unless one uses enormous size of beam. In a sense, we introduced a way to replace the shot-noise problem to the time-delay-noise problem. Besides, time-delay noise is different from clock noise that is introduced in \cite{DFI1} as fundamental noise in a couple of senses: (i) only a differential mode of the fluctuation between two beams matters (ii) noise disappears if two beams share the same time delay. We hope invention of a time-delay device that has no fluctuation in the observation band in the future.

On the other hand, we tried a configuration to cancel time-delay noise: by putting the device into the two arms of a Michelson interferometer. In such schemes, beamsplitter motions in these Michelson interferometers impose noise, which must be cancelled by introducing another set of  DFI detector. However it turned out that the cancellation of time-delay noise makes the sensitivity worse than that of a DFI without time delay.

In the paper we have focused on a ground-based detector, but the DFI is useful also in space-based detectors~\cite{LISA}\cite{DECIGO}. Their sensitivity will be limited by shot noise at high frequencies and by acceleration noise that shows $f^{-2}$ curve at low frequencies. Acceleration noise, which couples through motion of mirrors, will not appear in a DFI detector. While shot noise of a conventional DFI without time delay that shows $f^{-2}$ curve has a low advantage over removing acceleration, shot noise of time-delay DFI that shows $f^{-1}$ curve can be a remarkable improvement. Here a question is if we can make the time delay of $\sim100$~mHz necessary for space-based detectors, but we may expect improvements. Time-delay noise is also a big problem.

\ack
We wish to thank Seiji Kawamura, Archana Pai, Rick Savage, Malik Rakhmanov and Peter Beyersdorf for valuable discussions. Research of K.S. and Y.C. are supported by the Alexander von Humboldt Foundation. K.G. is supported by the U.S. National Science Foundation under Cooperative Agreement No. PHY-0107417 and PHY-0457264.

\appendix
\section{EIT Noise}\label{app}

A time-delay device with EIT is realized by a cell filled by vapor atoms and pump field injected to the cell in addition to the probe light that measures gravitational waves. While the cell does not transmit the probe light without the pump field, a transition of the atomic state due to the interaction with the pump field makes the cell transparent to the light at a particular frequency. The transition makes a refractive index of the cell rapidly vary and slows group velocity of the light, which results in the time-delay effect.~\cite{EIT} Fluctuation of the time delay, or EIT noise, comes from thermal motion of atoms. The probe light will see different numbers of atoms at different time. The time delay is proportional to the number of atoms that interact with the probe light, so the number fluctuation makes the time-delay fluctuation. In the case we have two counter-propagating beams to one EIT medium, where only a differential mode of the fluctuations matters as we will subtract the outputs to obtain the signal of DFI, it is not only the self-correlation function but also the cross-correlation function between two beams that should be taken into account. As a result, the power spectrum density of the differential fluctuation is given as
\begin{eqnarray}
S^\mathrm{diff}_\tau(\Omega)\simeq\frac{T_0^2}{N}\tau_Me^{-\Omega\tau_M}\left(1-\frac{v_T}{L\Omega}\sin{\left[\frac{L\Omega}{v_T}\right]}\right).\label{eq:final}
\end{eqnarray}
Here $T_0=L/v_g$ is the time delay with $L$, a length of the medium, and $v_g$, averaged group velocity of the light, $N$ is the {\it effective} number of atoms in the Gaussian beam, and $\tau_M=\sqrt{2}w/v_T$ is averaged time that an atom stays in the beam according to the Maxwell distribution ($w$ is a beam radius and $v_T$ is averaged speed of the atoms). Let us put the time delay of $T_0=0.02$~s. Typical parameters; $L=10$~cm, $w=1$~cm, $v_T=300$~m/s, and $N=10^{14}$, which corresponds to the atom density of $10^{-12}\ \mathrm{cm}^{-3}$, give us the EIT-noise level at 50~Hz, for example, as $\sqrt{S_\tau}\sim 3\times10^{-13}\ \mathrm{s}/\sqrt{\mathrm{Hz}}$. In displacement, this is $\sim100\ \mu\mathrm{m}/\sqrt{\mathrm{Hz}}$, which is much larger than other noise in gravitational-wave detectors. The noise level increases almost linearly up to $f\sim2$~kHz, and exponentially decreases from $f\sim1/(2\pi\tau_M)\sim3$~kHz. One could make this decay start from frequencies as low as 50~Hz, but it requires the temperature as low as 80~mK or the beam radius as large as 60~cm, which would be practically challenging. A use of Michelson interferometer to remove EIT noise, as we show in this paper, will be necessary.


\begin{thebibliography}{99}
\bibitem{Sagnac} A difference in the responses to gravitational waves and to displacement noise in the Sagnac interferometer is well known, but we do not know who pointed out this for the first time.
\end{thebibliography}
\end{document}